\documentclass{conm-p-l}

\copyrightinfo{2006}{}

\usepackage{amsmath,amssymb}

\usepackage{graphicx}

\usepackage{subfigure}

\newcommand{\La}{\Lambda}

\newcommand{\sg}{\sigma}

\newcommand{\ra}{\rightarrow}
\newcommand{\dl}{\delta}

\begin{document}

\title[Segment Description of Chaos]
{A Markov Chain Approximation of a Segment Description of Chaos}

\author{Alexander Labovsky}

\address{Department of Mathematics, University of Missouri, 
Columbia, MO 65211}

\email{alabovsky@sc.fsu.edu}

\thanks{Labovsky's current address: Florida State University,
Department of Scientific Computing,
Tallahassee, FL 32306-4120.}

\author{Y. Charles Li}

\address{Department of Mathematics, University of Missouri, 
Columbia, MO 65211}

\curraddr{}
\email{liyan@missouri.edu}

\subjclass{Primary 37, 76; Secondary 34, 35}
\date{}


\dedicatory{}

\begin{abstract}
We test a Markov chain approximation to the segment description (Li, 2007) 
of chaos (and turbulence) on a tent map, the Minea system, the H\'enon map,
and the Lorenz system. For the tent map, we compute the probability transition 
matrix of the Markov chain on the segments for segment time length 
(iterations) $T = 1, 2, 3, 100$. The matrix has $1, 2, 4$ tents corresponding 
to $T = 1, 2, 3$; and is almost uniform for $T = 100$. As $T \ra +\infty$, 
our conjecture is that the matrix will approach a uniform matrix (i.e. every 
entry is the same). For the simple fixed point attractor in the 
Minea system, the Reynolds average performs excellently and better than 
the maximal probability Markov chain and segment linking. But for the strange 
attractors in 
the H\'enon map, and the Lorenz system, the Reynolds average performs very
poorly and worse than the maximal probability Markov chain and segment linking.
\end{abstract}

\maketitle

\tableofcontents

\section{Markov Chain Approximation}

In \cite{Li07}, we showed that any orbit inside an attractor (chaotic or turbulent attractors are the most interesting ones) can be uniformly approximated on the infinite time interval $t \in [0, \infty )$ by an infinite sequence of segments out of a finite number of segments. The attractor (or its attracting neighborhood) is partitioned into $N$ small neighborhoods $\{ A_n \}_{n=1,2, \cdots , N}$. Each neighborhood $A_n$ is attached with an orbit segment $s_n$ over a fixed time interval $t \in [0,T]$. Denote by $F^t$ the evolution operator. The flow tube $\cup_{t \in [0,T]} F^t(A_n)$ is approximated by the segment $s_n$. If $F^T(A_n) \cap A_m \neq \emptyset$, then the flow tube $\cup_{t \in [T,2T]} F^t(F^T(A_n) \cap A_m)$  is approximated by the segment $s_m$. This process can be continued to $t \ra \infty$. If we re-distribute the Lebesgue measure of $F^T(A_n) \cap A_m$ uniformly on $A_m$ (i.e. we assume the points in $F^T(A_n) \cap A_m$ as random points with uniform probability distribution in $A_m$), then we obtain a Markov chain approximation. When $A_n$'s are small and the attractor has good mixing properties, we expect this Markov chain approximation to perform very well. The key element for this Markov chain approximation is the transition matrix \cite{KS60}, \cite{DLZ01}
\begin{equation}
\La = \left ( \rho_{ij} \right ) \ ,
\label{mctm}
\end{equation}
where
\[
\rho_{ij} = \frac{ \mu (F^T(A_j) \cap A_i)}{\mu (F^T(A_j))} \ ,
\]
and $\mu$ is the Lebesgue measure. We can simply label $A_n$ and $s_n$ by the letter $n$. Then our Markov chain is defined by the transition matrix $\La$ acting on $N$ letters $\{ 1,2, \cdots , N\}$.

\section{Ulam Approximation}

The Transfer Operator (Perron-Frobenius Operator) is defined as,
\begin{equation}
\nu_{n+1}(A) = \nu_n(F^{-T}(A))=\sum_{j=1}^{N} \nu_n(F^{-T}(A) \cap A_j) \  ,
\label{to}
\end{equation}
where $\nu_n$'s are measures. The Ulam approximation of the transfer operator is defined as,
\begin{equation}
\nu_{n+1}(A) =\sum_{j=1}^{N} \frac{\mu (F^{-T}(A) \cap A_j)} { \mu (A_j)} \nu_n (A_j) \  ,
\label{ua}
\end{equation}
where $\mu$ is the Lebesgue measure.
If we re-distribute $\nu_n(F^{-T}(A) \cap A_j)$ uniformly on $A_j$ (i.e. we assume the points in $F^T(A) \cap A_j$ as random points with uniform probability distribution in $A_j$), then we obtain the Ulam approximation by
\[
\nu_n(F^{-T}(A) \cap A_j)  \sim \frac{\mu (F^{-T}(A) \cap A_j)} { \mu (A_j)} \nu_n (A_j) \  .
\]
The transition matrix of the Ulam approximation is
\begin{equation}
P = \left ( \sg_{ij} \right ) \ ,
\label{uatm}
\end{equation}
where
\[
\sg_{ij} = \frac{ \mu (F^{-T}(A_i) \cap A_j)}{\mu (A_j)} \  .
\]

If $F^T$ is $1-1$ and we replace the partition $\{ A_i \}$ by $\{ F^T(A_i)\}$, then
\[
\sg_{ij} = \frac{ \mu (A_i \cap F^T(A_j))}{\mu (F^T(A_j))}  = \rho_{ij}  \ .
\]

\section{Asymptotic Cycles \label{SAC}}

Using the segments $\{ s_n \}_{n=1,2, \cdots , N}$, one can identify certain 
pseudo-orbits (called segment linking orbits) and the associated sequence of $A_n$'s:
\begin{eqnarray*}
& & \cdots s_{n_{-2}} s_{n_{-1}} s_{n_{0}} s_{n_{1}} s_{n_{2}} \cdots \\
& & \cdots A_{n_{-2}} A_{n_{-1}} A_{n_{0}} A_{n_{1}} A_{n_{2}} \cdots
\end{eqnarray*}
where $s_{n_{j}}$ attaches to $A_{n_{j}}$ and the end point of $s_{n_{j}}$ 
belongs to $A_{n_{j+1}}$ (recall also that the starting point of 
$s_{n_{j}}$ belongs to $A_{n_{j}}$). Since there are only finite $s_n$'s; 
as $n_{j} \ra +\infty$, some $s_{n_{j}}$ has to repeat itself, then all the segments 
after it repeat themselves too. Thus all these pseudo-orbits are always asymptotic 
to cycles:
\[
\cdots \cdots s_{n_{j_1}} \cdots s_{n_{j_k}} s_{n_{j_1}} \cdots s_{n_{j_k}} \cdots
\]
where $k$ can be $1$ in which case the asymptotic cycle is a ``fixed point''.

\section{Transition Matrix}

\begin{figure}[!h]
\centering
\subfigure[$T=1$]{\includegraphics[width=2.4in,height=3.0in]{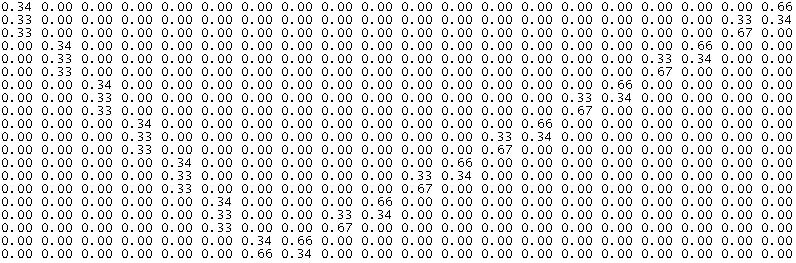}}
\subfigure[$T=2$]{\includegraphics[width=2.4in,height=3.0in]{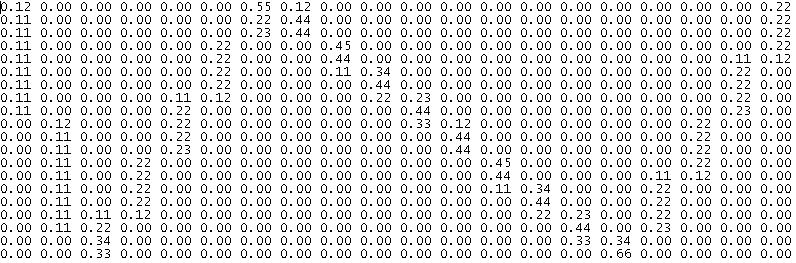}}
\subfigure[$T=3$]{\includegraphics[width=2.4in,height=3.0in]{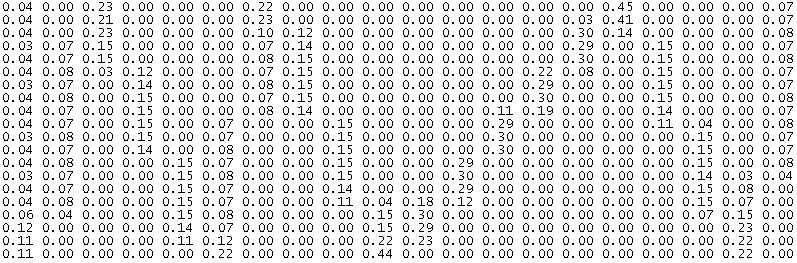}}
\subfigure[$T=100$]{\includegraphics[width=2.4in,height=3.0in]{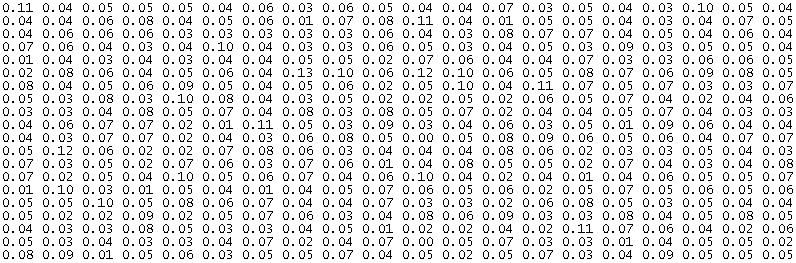}}
\caption{Transition matrices for the tent map}
\label{tm}
\end{figure}

We shall compute the transition matrix for the tent map:
\[
x_{n+1} = f(x_n), \ x_n \in [0,1]; 
\]
where
\[
 f(x_n) = \frac{x_n}{a} \ (x_n<a),
\ f(x_n) = \frac{x_n-1}{a-1} \ (x_n \geq a); 
\]
and $a$ is a parameter $a\in (0,1)$. We choose $a=1/3$. We break the interval 
$[0,1]$ into $20$ subintervals. Take $100$ initial conditions in each subinterval. 
For each initial condition $x_0$ in the $i$-th subinterval, we compute $x_T$ as the 
$T$-iteration starting from $x_0$. If $x_T$ belongs to $j$-th subinterval, 
we increase the $(j,i)$-th entry of the transition matrix by $\frac{1}{100}$. 
Initially all the entries of the transition matrix are set to $0$. Since the 
iterated map $f^T$ has $2^{T-1}$ tents, the transition matrix also resembles 
this pattern. Our conjecture is that when $T \ra + \infty$, the transition matrix
approaches a uniform matrix with every entry being $1/20$. See Figure \ref{tm} for 
an illustration.

\section{Different Types of Orbits}

We shall compute several types of orbits:
\begin{enumerate}
\item {\bf The true orbit}. It is computed by Runge-Kutta method for 
differential equations, and by iterations for maps; for a time interval 
$t \in [0, T^*]$. 
\item {\bf The segment orbit}. First we cut the total time interval $[0, T^*]$
into subintervals of length $T$. The interested region in the phase space, where 
the attractor lives is partitioned into $N$ small neighborhood 
$\{ A_n \}_{n=1,2, \cdots , N}$ (usually squares or cubes). If at time $t = j T$, the 
true orbit lands in some $A_n$, then $(j+1)$-th segment will be the one attached to 
$A_n$ (usually this segment is computed with the initial condition at the center of 
the square or cube for a time length of $T$). For $j = 0, 1, 2, \cdots$; all the segments
together form the segment orbit. It is a uniform approximation of the 
true orbit for all positive time. 
\item {\bf The segment linking orbit}. If the initial condition of the true orbit
lands in some $A_n$, then the first segment will be the one attached to $A_n$. If the 
first segment ends in some $A_m$, then the second segment will be the one attached to 
$A_m$. By repeating this process, we generate the segment linking orbit. As discussed
in Section \ref{SAC}, the segment linking orbit is asymptotic to some cycle. The 
segment orbit uniformly approximates a particular orbit in the basin of attraction,
while the segment linking orbit only uses the initial condition of the true orbit, and 
is not a uniform approximation of the true orbit. In reality, the segment orbit is 
difficult to obtain without the full knowledge of the true orbit, while the 
segment linking orbit is trivial to obtain once all the segments attached to the $A_n$'s are known. 
\item {\bf The maximal probability Markov chain}. If the initial condition of the 
true orbit lands in some $A_n$, then the first segment will be the one attached to $A_n$.
Then we test many random initial points in $A_n$, see where they land at time $t=T$,
and we pick maximal probability partition element $A_m$. The second segment will 
be the one attached to $A_m$. Repeating the process, we generate the maximal 
probability Markov chain.
\item {\bf The Reynolds average orbit}. We pick initial conditions near the 
initial condition of the true orbit, compute the orbit corresponding to 
each initial condition, at each $t \geq 0$, we do an algebraic average of all the 
orbits, then we get the Reynolds average orbit.
\end{enumerate}

Overall, the segment orbit is a uniform approximation of a particular orbit, while 
the segment linking orbit, the maximal probability Markov chain and the Reynolds 
average orbit are all aiming at certain average property of the attractor.

\section{Minea System}

\begin{figure}[!h]
\centering
\subfigure[true orbit]{\includegraphics[width=2.3in,height=2.3in]{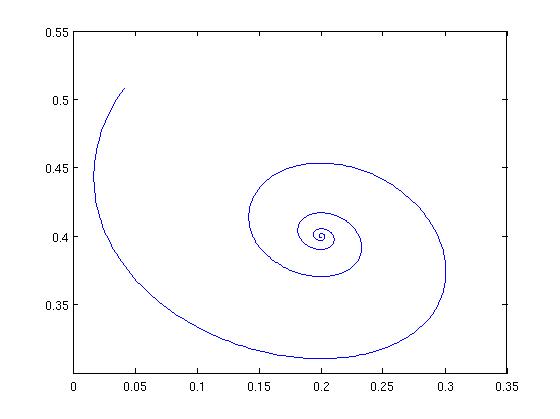}}
\subfigure[segment orbit]{\includegraphics[width=2.3in,height=2.3in]{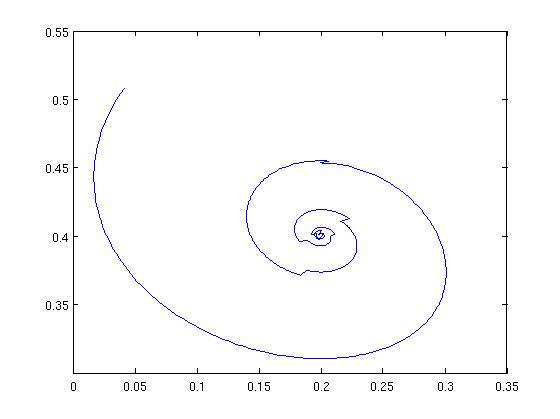}}
\subfigure[segment linking orbit]{\includegraphics[width=2.3in,height=2.3in]{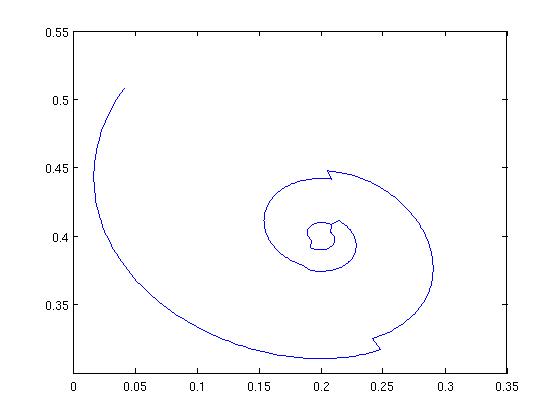}}
\subfigure[maximal probability Markov chain]{\includegraphics[width=2.3in,height=2.3in]{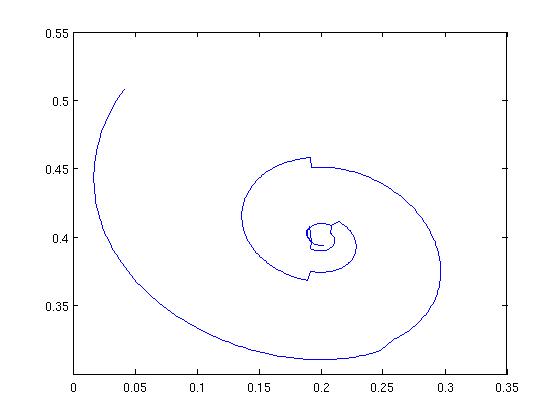}}
\subfigure[Reynolds average orbit]{\includegraphics[width=2.3in,height=2.3in]{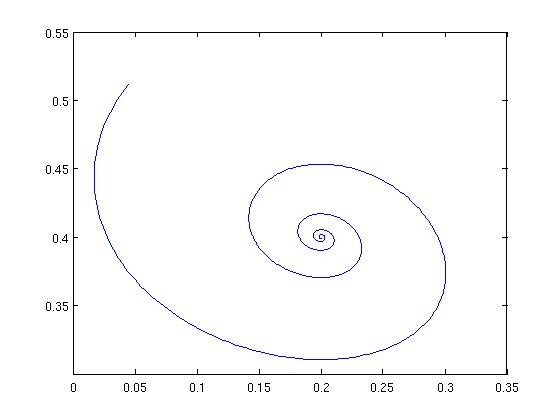}}
\caption{Different types of orbits for the Minea system.}
\label{minea}
\end{figure}

Consider the Minea system
\begin{eqnarray*} 
\frac{d u_1}{dt} &=& 1-u_1-\delta u_2^2 \ ,\\
\frac{d u_2}{dt} &=& \delta u_1 u_2-u_2\ ;
\end{eqnarray*}
where $(u_1, u_2) \in (0,1)\times(0,1)$ and $\dl$ is a parameter. 
When $\delta > 1$, all the orbits converge to one point \cite{T88}. 
We choose $\delta = 5$ for our simulations here. The total simulation time $T^* = 10$, 
while the segment time $T=1$ (i.e. $10$ segments along each orbit). The interested 
region in the phase plane is $(u_1, u_2) \in (0,1)\times(0,1)$. We cut this unit 
square into $60 \times 60$ subsquares (i.e. the $A_n$'s). The segment attached to 
each $A_n$ starts from the center of $A_n$. The true orbit starts from the initial 
point ($\frac{1}{24}, 0.5 + \frac{1}{120}$). For the calculation of the maximal 
probability Markov chain, we test $1000$ random initial points in $A_n$. For the 
calculation of the Reynolds average orbit, we pick $4$ initial conditions within 
$\frac{1}{120}$ distance from the initial condition of the true orbit. The results 
are shown in Figure \ref{minea}. As expected, the Reynolds average 
orbit is a better approximation of the true orbit than the other types of orbits.

\section{H\'enon Map}

Consider the H\'enon map
\begin{eqnarray*}
x_{n+1}&=&y_n+1-1.4x_n^2,\\
y_{n+1}&=&0.3x_n.
\end{eqnarray*}

\begin{figure}[!h]
\centering
\subfigure[true orbit]{\includegraphics[width=2.3in,height=2.3in]{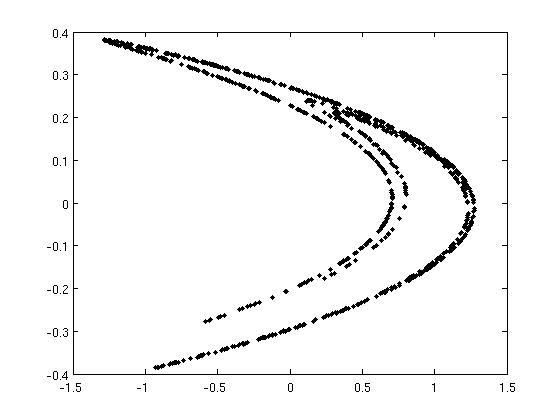}}
\subfigure[segment orbit]{\includegraphics[width=2.3in,height=2.3in]{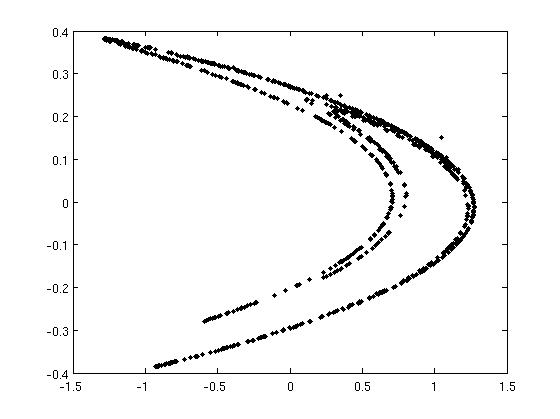}}
\subfigure[segment linking orbit]{\includegraphics[width=2.3in,height=2.3in]{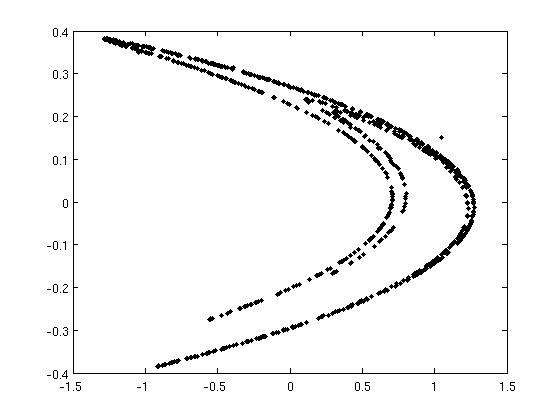}}
\subfigure[maximal probability Markov chain]{\includegraphics[width=2.3in,height=2.3in]{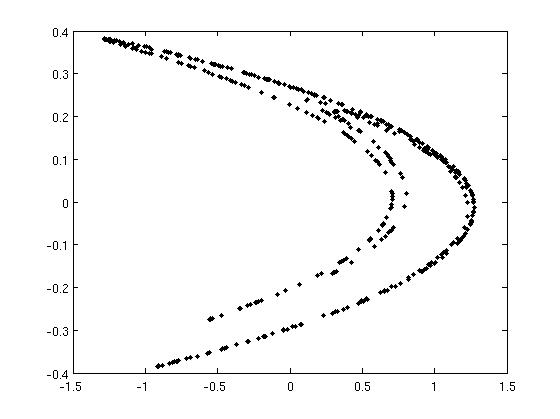}}
\subfigure[Reynolds average orbit (of 2 orbits)]{\includegraphics[width=2.3in,height=2.3in]{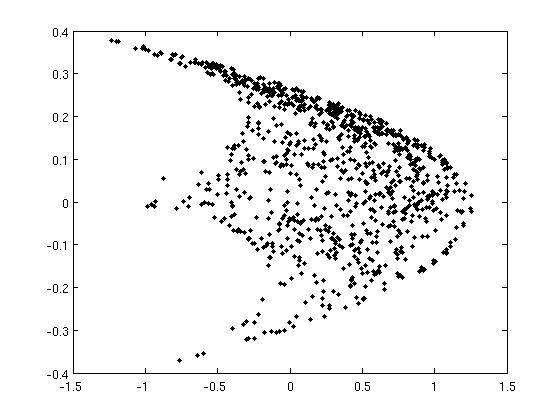}}
\subfigure[Reynolds average orbit (of 4 orbits)]{\includegraphics[width=2.3in,height=2.3in]{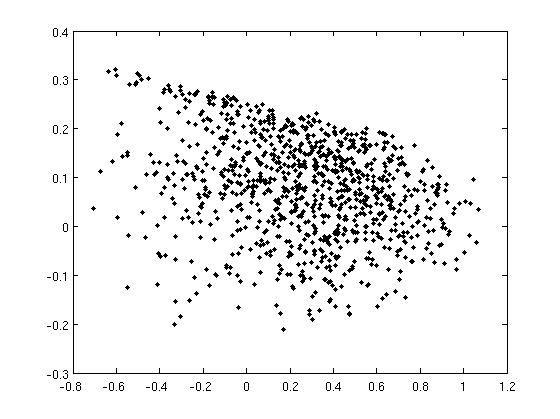}}
\caption{Different types of orbits for the H\'enon map.}
\label{henon}
\end{figure}

The total simulation time $T^* = 1000$ (iterations), 
while the segment time $T=250$ (i.e. $4$ segments along each orbit). The interested 
region in the phase plane is $(x_n, y_n) \in (-2,2)\times(-2,2)$. We cut this 
square into $40 \times 40$ subsquares (i.e. the $A_n$'s). The segments attached to 
each $A_n$ starts from the center of $A_n$. The true orbit starts from the initial 
point ($0.631354477,0.189406343$). For the calculation of the maximal 
probability Markov chain, we test $1000$ random initial points in $A_n$. For the 
calculation of the Reynolds average orbit, we pick $2$ or $4$ initial conditions within 
$10^{-6}$ distance from the initial condition of the true orbit. The results 
are shown in Figure \ref{henon}. One thing is clear, that is, the 
Reynolds average orbits are definitely bad approximations to the true orbit, and 
bad descriptions of the strange attractor.

\section{Lorenz System}

Consider the Lorenz system 
\begin{eqnarray*}
\frac{dx}{dt} &=& a(y-x),\\
\frac{dy}{dt} &=& x(b-25 z)-y,\\
\frac{dz}{dt} &=& 25 xy-cz,
\end{eqnarray*}
where we choose $a=10$, $b=28$, $c=8/3$.

\begin{figure}[!h]
\centering
\subfigure[true orbit]{\includegraphics[width=2.3in,height=2.3in]{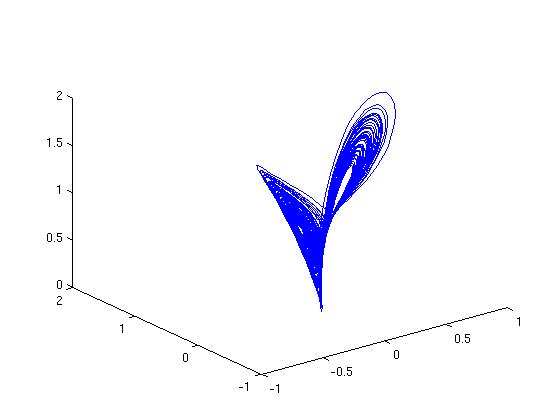}}
\subfigure[segment orbit (10)]{\includegraphics[width=2.3in,height=2.3in]{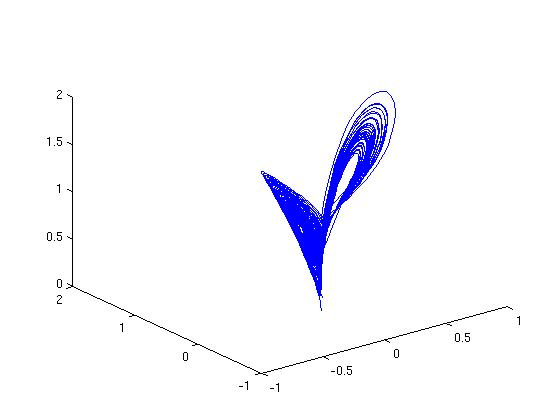}}
\subfigure[segment orbit (100)]{\includegraphics[width=2.3in,height=2.3in]{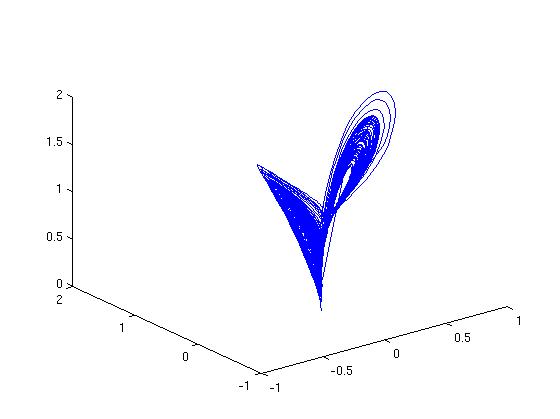}}
\subfigure[segment linking orbit (10)]{\includegraphics[width=2.3in,height=2.3in]{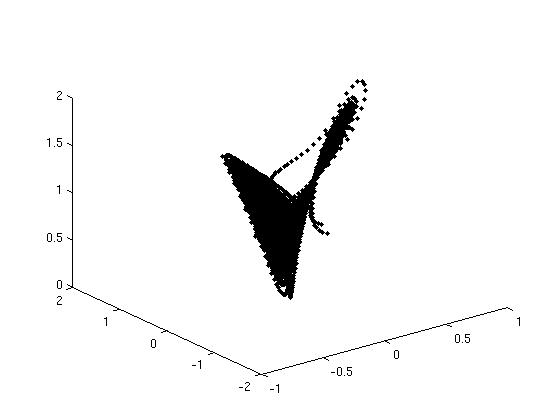}}
\subfigure[segment linking orbit (100)]{\includegraphics[width=2.3in,height=2.3in]{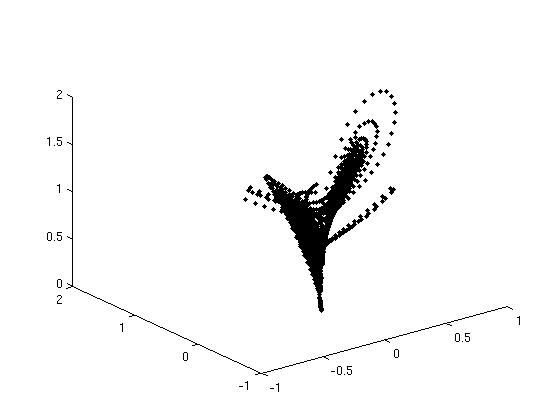}}
\caption{Different types of orbits for the Lorenz system (part 1).}
\label{lorenz1}
\end{figure}

\begin{figure}[!h]
\centering
\subfigure[maximal probability Markov chain (10)]{\includegraphics[width=2.3in,height=2.3in]{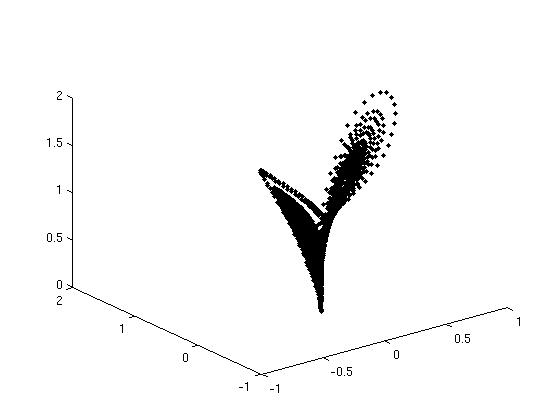}}
\subfigure[maximal probability Markov chain (100)]{\includegraphics[width=2.3in,height=2.3in]{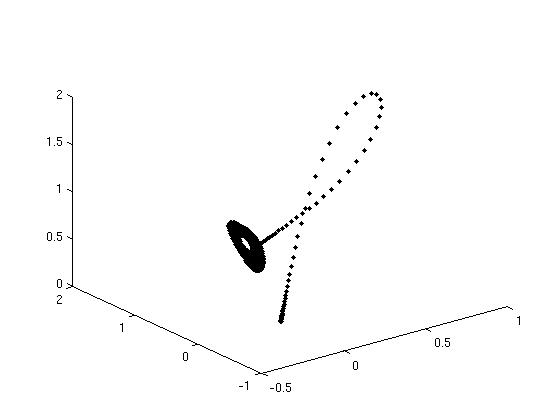}}
\subfigure[Reynolds average orbit (of 2 orbits)]{\includegraphics[width=2.3in,height=2.3in]{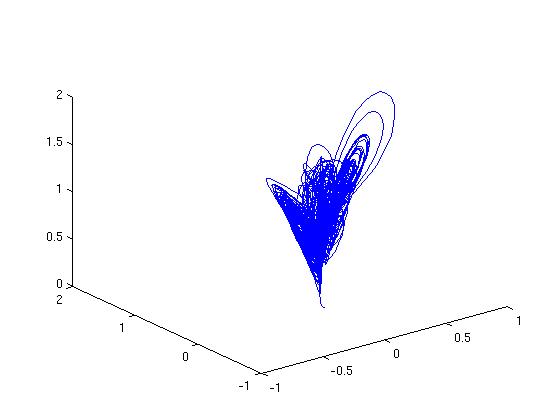}}
\subfigure[Reynolds average orbit (of 4 orbits)]{\includegraphics[width=2.3in,height=2.3in]{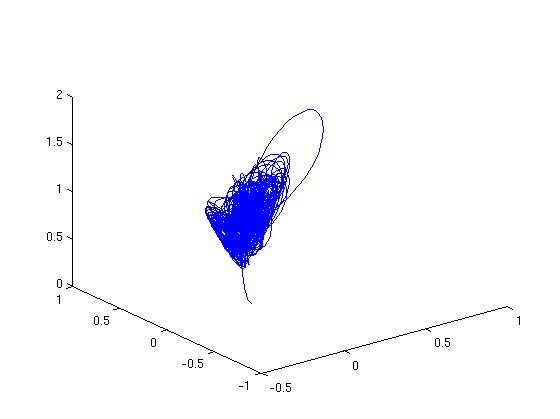}}
\caption{Different types of orbits for the Lorenz system (part 2).}
\label{lorenz2}
\end{figure}

The total simulation time $T^* = 100$, 
while the segment time is either $T=10$ (i.e. $10$ segments along each orbit) or 
$T=1$ (i.e. $100$ segments along each orbit). The interested 
region in the phase space is $(x, y, z) \in (-0.8,0.8)\times(-1,1)\times(0,2)$. We 
cut this region into $16 \times 16 \times 16$ subregions (i.e. the $A_n$'s). The 
segments attached to 
each $A_n$ starts from the center of $A_n$. The true orbit starts from the initial 
point ($0.1, 0, 0$). For the calculation of the maximal 
probability Markov chain, we test $1000$ random initial points in $A_n$. For the 
calculation of the Reynolds average orbit, we pick $2$ or $4$ initial conditions within 
$0.05$ distance from the initial condition of the true orbit. The results 
are shown in Figures \ref{lorenz1} and \ref{lorenz2}. Once again the 
Reynolds average orbits are definitely bad approximations to the true orbit, and 
bad descriptions of the strange attractor. Naturally as the number of the segments along
an orbit increases, the segment linking orbit and the maximal probability Markov chain
deviate further away from the true orbit.

\section{Conclusion and Discussion}

As discussed in details in \cite{Li07} \cite{Li07b} \cite{Li07c}, an effective description of 
turbulence means a solution to the problem of turbulence. An effective description of chaos 
is also very useful in applications of chaos theory. 

It is dangerous to draw strong conclusion with limited numerical experiments. But it is clear 
from these numerical experiments here that the Reynolds average is a terrible description of 
strange attractors. The segment description has the potential to generate useful descriptions of 
strange attractors. Both the segment linking and the Markov chain have the potential 
to generate useful characterizations on certain average properties of the strange attractors. 

Traditionally the Reynolds average is based upon the setup of typical orbits being high 
frequency oscillations around a mean orbit. This is not the case inside a strange attractor. 
Therefore, one should not expect that the Reynolds average can be a good description on any sort 
of average property of the strange attractor.  For the segment description, computing the 
segments can be costly if the partition is fine enough. Computing the segment orbit associated 
with a true orbit needs the full knowledge of the true orbit. So the potential of the segment 
description lies at the statistics of the segments rather than any individual segment orbit. 
Segment linking is a convenient way to 
start the study on the statistics of the segments. Markov chain is a first level study on 
the statistics of the segments. Practically it is very costly to generate the probability 
transition matrix for the Markov chain.

Finally what kind of average properties that a strange attractor may possess are not clear
at all. If the strange attractor has ergodicity, then tracking one orbit is enough to get 
a good picture of the strange attractor. On the other hand, especially in higher dimensions, 
often the chaotic (turbulent) dynamics is transient. What kind of average properties that 
such transient dynamics may possess are even less clear.

\end{document}